\documentclass[aps,showpacs,prb,reprint,longbibliography,superscriptaddress]{revtex4-1}

\usepackage[colorlinks=true,linkcolor=blue,citecolor=blue,urlcolor=blue,bookmarks=false]{hyperref}

\usepackage{setspace} 
\usepackage{graphicx}
\usepackage{amsmath}
\usepackage{color}
\usepackage{amsmath}
\usepackage{amssymb}
\usepackage{verbatim}
\usepackage{latexsym}
\usepackage{enumerate} 
\usepackage{bm} 
\usepackage[caption=false]{subfig}
\begin{document}

\title{Phonon-assisted optical absorption in BaSnO$_3$ from first principles}

\author{Bartomeu Monserrat} \email{bm418@cam.ac.uk}
\affiliation{Department of Physics and Astronomy, Rutgers University,
  Piscataway, New Jersey 08854-8019, USA} 
\affiliation{TCM Group,
  Cavendish Laboratory, University of Cambridge, J.\ J.\ Thomson
  Avenue, Cambridge CB3 0HE, United Kingdom}
\author{Cyrus E. Dreyer} 
\affiliation{Department of Physics and Astronomy, Rutgers University,
  Piscataway, New Jersey 08854-8019, USA} 
\author{Karin M. Rabe} 
\affiliation{Department of Physics and Astronomy, Rutgers University,
  Piscataway, New Jersey 08854-8019, USA}

\date{\today}

\begin{abstract} 
The perovskite BaSnO$_3$ provides a promising platform for the realization of an earth abundant $n$-type transparent conductor. Its optical properties are dominated by a dispersive conduction band of Sn $5s$ states, and by a flatter valence band of O $2p$ states, with an overall indirect gap of about $2.9$\,eV. 
Using first-principles methods, we study the optical properties of BaSnO$_3$ and show that both electron-phonon interactions and exact exchange, included using a hybrid functional, are necessary to obtain a qualitatively correct description of optical absorption in this material. In particular, the electron-phonon interaction drives phonon-assisted optical absorption across the minimum indirect gap and therefore determines the absorption onset, and it also leads to the temperature dependence of the absorption spectrum. Electronic correlations beyond semilocal density functional theory are key to detemine the dynamical stability of the cubic perovskite structure, as well as the correct energies of the conduction bands that dominate absorption. Our work demonstrates that phonon-mediated absorption processes should be included in the design of novel transparent conductor materials.
\end{abstract}

\maketitle

\section{Introduction}

Transparent conductors are materials which simultaneously exhibit optical transparency in the visible and high carrier mobilities.~\cite{tco_review_2000,tco_review_2005,tco_solar_cell_review_2007,tco_review_2012,tco_review_ptype_2016,metal_oxide_review_2016} As such, they are of interest for optoelectronic applications like photovoltaics and flat panel displays. Commercially available transparent conductors are mostly based on In$_2$O$_3$, a large band gap semiconductor, that exhibits high conductivities when doped with tin. However, the scarcity of indium and the associated high costs have fuelled a significant research effort to find alternative materials that could act as transparent conductors.

Several strategies for designing novel transparent conductors have been pursued. The most widespread route is to replace In$_2$O$_3$ with alternative large band gap oxide semiconductors, and in this paper we focus on one of the most promising examples in this area, BaSnO$_3$.~\cite{bso_early_exp_cheong,bso_original_report,bso_exp_characterization} Alternative routes include the use of non-oxide semiconductors,~\cite{tc_ternary_zunger} graphene,~\cite{tco_graphene} metal nanowires,~\cite{tco_metal_nanowires}, correlated metals,~\cite{tco_correlated_metal} or band engineering of the bulk.~\cite{zunger_bulk_band_engineering}

The cubic perovskite BaSnO$_3$ has emerged as a promising candidate for transparent conductor applications due to the high $n$-type mobilities exhibited when doped with lanthanum, while retaining its favourable optical properties.~\cite{bso_early_exp_cheong,bso_original_report,bso_exp_characterization} The high mobility is belived to arise from the Sn $5s$ character of the conduction band, which provides a small effective mass~\cite{bso_original_report,bso_exp_effective_mass} and a low density of states reducing the phase space available for electron scattering.~\cite{bso_transport_walle} The optical properties are dominated by an indirect gap that marks the absorption onset at $2.9$\,eV, and a strong optical absorption starting at $3.1$~eV and marking the direct gap.~\cite{bso_exp_characterization} Thin films of doped samples exhibit transmittances of about $80$\% in the visible range.~\cite{bso_early_exp_chen,bso_exp_characterization}

While first-principles methods have been used extensively to study the structure,~\cite{bso_hse_bands_and_phonons} dynamics,~\cite{bso_hse_bands_and_phonons,bso_phonons_bands,bso_hybrid_functional} band structure,~\cite{bso_hse_bands_and_phonons,bso_phonons_bands,bso_hybrid_functional} transport,~\cite{bso_transport_walle} and doping~\cite{bso_defects_scanlon} of BaSnO$_3$, a full characterization of the optical properties is missing. This is because optical absorption in indirect gap semiconductors is mediated by lattice vibrations and therefore a full description of the absorption onset in BaSnO$_3$ requires the inclusion of electron-phonon interactions. Although the theory of phonon-assisted optical absorption has been known for a long time,~\cite{williams_phonon_assisted_optics,lax_phonon_assisted_optics,phonon_assisted_abs_hbb} it has only recently been combined with first-principles methods to study absorption in indirect gap materials~\cite{phonon_assisted_abs_cohen,giustino_nat_comm,patrick_elph_long,phonon_assisted_abs_stochastic,phonon_assisted_abs_one_shot} and free-carrier absorption in doped semiconductors.~\cite{kioupakis_free_carrier_abs,kioupakis_sno2_free_carrier_abs,kioupakis_sno2_free_carrier_abs2}

In this paper, we use first principles methods to study the optical absorption spectrum of BaSnO$_3$ including the effects of lattice vibrations. We find that if phonon mediated processes are included, then (i) the absorption onset of BaSnO$_3$ coincides with the indirect band gap, (ii) the indirect absorption onset is redshifted by $0.1$\,eV in going from $0$\,K to $300$\,K, and (iii) absorption below $7$\,eV, corresponding to the highly dispersive conduction Sn $5s$ band, is noticeably enhanced compared to the static lattice counterpart. 
We have performed calculations using both semilocal and hybrid exchange correlation functionals, and find that the lattice stability, phonon dispersion, electronic band structure, and absorption spectrum are highly sensitive to the choice of functional, with the hybrid functional providing the best agreement with available experimental data. Overall, our results demonstrate that an accurate description of the optical properties of BaSnO$_3$ requires the inclusion of both electron-phonon interactions and electron-electron interactions beyond semilocal density functional theory. 

The rest of the paper is organized as follows. In Sec.~\ref{sec:equilibrium} we present the calculated equilibrium properties of BaSnO$_3$ and in Sec.~\ref{sec:lattdyn} we describe our lattice dynamics results. In both sections we pay particular emphasis to the choice of exchange-correlation functional. In Sec.~\ref{sec:optics} we describe the phonon-assisted optical absorption formalism, and the results for the phonon-assisted indirect gap absorption as well as the temperature dependence of the absorption spectrum. We summarize our findings and reach our conclusions in Sec.~\ref{sec:conclusions}.

\section{Equilibrium properties} \label{sec:equilibrium}

\subsection{Computational details} \label{subsec:comput_details}

Our first principles calculations are performed using density functional theory (DFT)~\cite{PhysRev.136.B864,PhysRev.140.A1133} within the projector augmented-wave method~\cite{paw_original,paw_us_relation} as implemented in {\sc vasp}.~\cite{vasp1,vasp2,vasp3,vasp4} We use an energy cut-off of $400$\,eV and a BZ grid size of $8\times8\times8$ $\mathbf{k}$-points for the primitive cell, and commensurate grids for the supercells. We test four distinct exchange correlation functionals for the calculation of the equilibrium volumes, electronic energy bands, and lattice dynamics, namely the local density approximation (LDA),~\cite{PhysRevLett.45.566,PhysRevB.23.5048,PhysRevB.45.13244} the generalized gradient approximation of Perdew-Burke-Ernzerhof (PBE),~\cite{PhysRevLett.77.3865} the PBE approximation for solids (PBEsol),~\cite{pbesol_functional} and the hybrid Heyd-Scuseria-Ernzerhof functional (HSE).~\cite{hse03_functional,hse06_functional,hse06_functional_erratum}

\subsection{Structure}

BaSnO$_3$ has the cubic perovskite structure of space group $Pm\overline{3}m$ with $5$ atoms in the primitive cell. We minimize the energy with respect to the cubic lattice parameter $a$ to find the equilibrium structure, and the results are presented in Table~\ref{tab:soft_mode}. The lattice parameter increases in the sequence of semilocal functionals LDA, PBEsol, and PBE, as expected. The hybrid HSE functional leads to a lattice parameter in close agreement with that of PBEsol, and overall the PBEsol and HSE lattice parameters agree best with experiment. 

\begin{table*}
  \setlength{\tabcolsep}{6pt} 
  \caption{Physical properties of BaSnO$_3$ using the LDA, PBEsol, PBE, and HSE exchange correlation functionals, and corresponding experimental data from Refs.~\onlinecite{bso_exp_sn_doping,bso_exp_characterization,bso_2004_overview,bso_exp_characterization,bso_exp_gap_thin_film,bso_exp_gap_thin_film2}. We show the cubic lattice parameter $a$, the minimum indirect gap $E_{\mathrm{g}}^{\mathrm{ind}}$ between the $R$ point and the $\Gamma$ point, the direct gap $E_{\mathrm{g}}^{\mathrm{direct}}$ at the $\Gamma$ point, and the energy of the $R$-point phonon mode labeled by the irreducible representation $R_5^{-}$, which is imaginary for the PBEsol and PBE functionals, indicating dynamical instability of the cubic phase in those cases.} 
  \label{tab:soft_mode}
  \begin{ruledtabular}
  \begin{tabular}{c|cccc}
            & $a$ (\AA)   & $E_{\mathrm{g}}^{\mathrm{ind}}$ (eV) & $E_{\mathrm{g}}^{\mathrm{direct}}$ (eV) &  $\omega_{R_5^{-}}$ (meV) \\ [0.1cm]
  \hline
  LDA        &  $4.07$ & $1.30$ & $1.83$ & $1.8$  \\
  PBEsol     &  $4.10$ & $1.02$ & $1.56$ & $3.2i$ \\
  PBE        &  $4.15$ & $0.66$ & $1.19$ & $3.3i$ \\
  HSE        &  $4.10$ & $2.73$ & $3.21$ & $11.3$ \\
  Experiment &  $4.12$~\cite{bso_exp_sn_doping} & $2.90$--$3.0$~\cite{bso_exp_characterization,bso_exp_gap_thin_film} & $3.10$--$3.60$~\cite{bso_2004_overview,bso_exp_characterization,bso_exp_gap_thin_film,bso_exp_gap_thin_film2} & -- \\
\end{tabular}
\end{ruledtabular}
\end{table*}

\subsection{Electronic band structure} \label{sec:bs}

\begin{figure}
\centering
\includegraphics[scale=0.35]{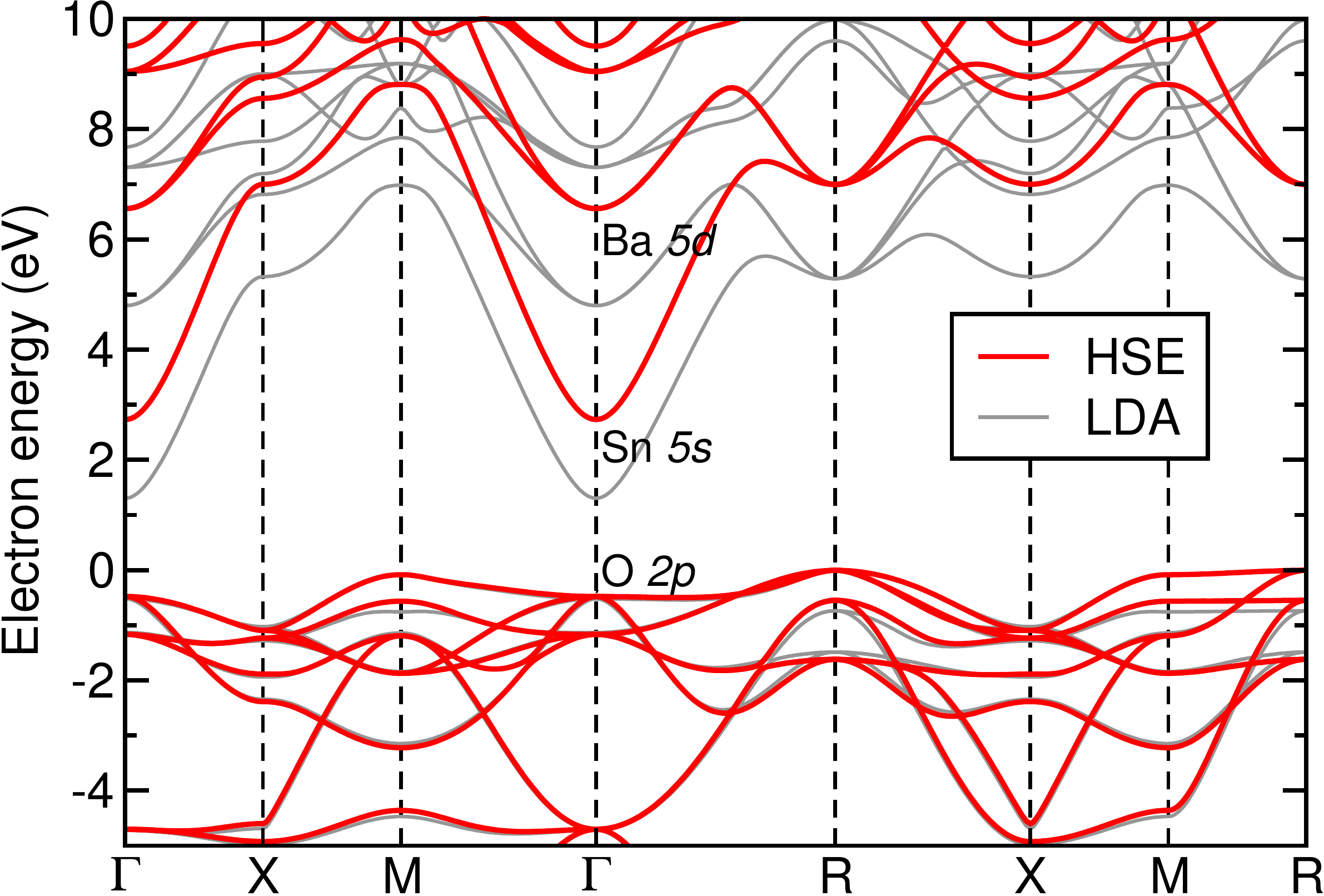}
\caption{Electronic band structure of BaSnO$_3$ at the LDA and HSE levels of theory. The state labels provide the dominant character of the bands around the band gap.} 
\label{fig:bs}
\end{figure}

We show the band structure of BaSnO$_3$ calculated using the LDA and HSE functionals in Fig.~\ref{fig:bs}. The valence bands are dominated by states of O $2p$ character, with the valence band maximum at the $R$ ($\frac{1}{2}$,$\frac{1}{2}$,$\frac{1}{2}$) point. The conduction band minimum is located at the $\Gamma$ point, with a band dominated by states of Sn $5s$ character that endow it with a strong dispersion and low effective mass. At about $4$\,eV above the minimum of the conduction band, we find bands dominated by Ba $5d$ states which exhibit a smaller dispersion. Our results agree with previous analysis.~\cite{bso_2004_overview}

The minimum band gap is indirect and has a value of $1.30$\,eV at the LDA level, and of $2.73$\,eV at the HSE level. By comparison, experimental estimates of the minimum gap are in the range $2.90$--$3.10$\,eV~\cite{bso_exp_characterization,bso_exp_gap_thin_film}. The minimum direct gap occurs at the $\Gamma$ point and has a value of $1.83$\,eV for LDA and $3.21$\,eV for HSE, with experimental estimates in the range $3.10$--$3.60$\,eV~\cite{bso_2004_overview,bso_exp_characterization,bso_exp_gap_thin_film,bso_exp_gap_thin_film2}. We note that the spread in experimental gaps might be related to the use of single crystals or thin films, with the latter providing larger estimates for the band gap sizes. We also note that our HSE band gaps are slightly larger than those previously reported.~\cite{bso_defects_scanlon,bso_transport_walle} We have observed that the band gap size at the HSE level is rather sensitive to the value of the lattice parameter, and we ascribe the difference with previous reports to the smaller lattice parameter in our calculations. We provide a list of the minimum indirect and direct gaps calculated with a range of exchange correlation functionals in Table~\ref{tab:soft_mode}.

It is important to note from Fig.~\ref{fig:bs} that the effect of including electron-electron correlations beyond semilocal DFT using the hybrid HSE functional is not limited to a rigid shift of the conduction bands. As an example, we consider the $\Gamma$ point gap between the O $2p$ valence band and the Sn $5s$ band at the bottom of the conduction band, denoted by $E_{\mathrm{g}}(\mathrm{O}2p\rightarrow\mathrm{Sn}5s)$ (the same as $E_{\mathrm{g}}^{\mathrm{direct}}$), and the $\Gamma$ point gap between the O $2p$ valence band and the Ba $5d$ band, denoted by $E_{\mathrm{g}}({\mathrm{O}2p\rightarrow\mathrm{Ba}5d})$. The $E_{\mathrm{g}}(\mathrm{O}2p\rightarrow\mathrm{Sn}5s)$ gap has values of $1.83$\,eV in LDA and $3.21$\,eV in HSE, leading to a shift of $1.38$\,eV. By comparison, the $E_{\mathrm{g}}({\mathrm{O}2p\rightarrow\mathrm{Ba}5d})$ gap has values of $5.32$\,eV in LDA and $7.04$\,eV in HSE, leading to a shift of $1.72$\,eV. Rather than only undergoing a rigid shift, the bands are also stretched when electron-electron correlations beyond semilocal DFT are included.

\section{Lattice dynamics} \label{sec:lattdyn}

\begin{figure}
\centering
\includegraphics[scale=0.35]{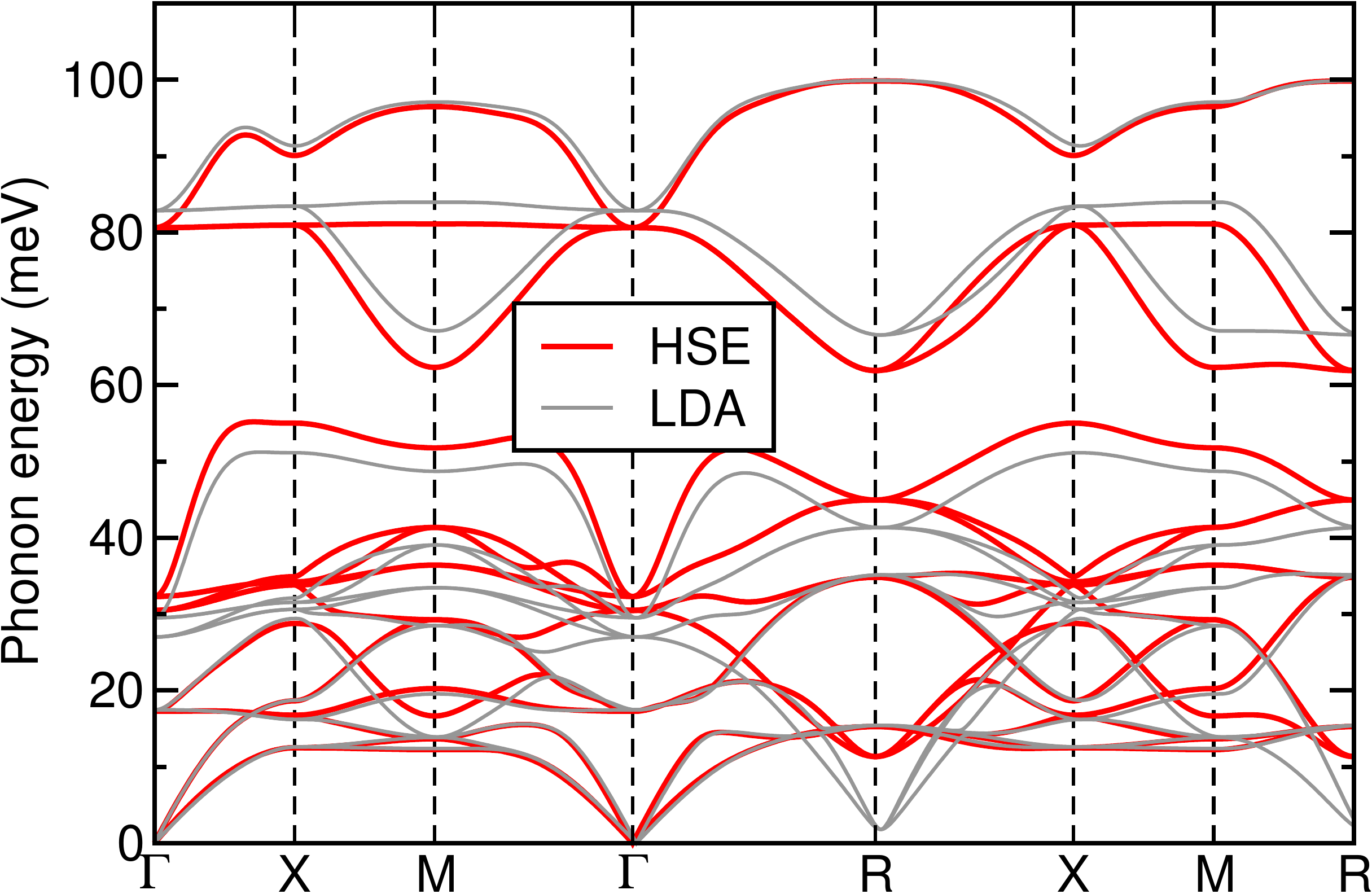}
\caption{Phonon dispersion of BaSnO$_3$ at the LDA and HSE levels of theory. The LO-TO splitting is neglected in this dispersion, but considered in Table~\ref{tab:lo-to}.} 
\label{fig:phonon}
\end{figure}

The lattice dynamics calculations have been performed using the same numerical parameters as those reported in Sec.~\ref{subsec:comput_details}, and the lattice parameters reported in Table~\ref{tab:soft_mode} corresponding to each functional used. We employ the finite displacement method~\cite{phonon_finite_displacement} in conjunction with nondiagonal supercells~\cite{non_diagonal} to construct the matrix of force constants, which is then Fourier transformed to the dynamical matrix and diagonalized to obtain the vibrational frequencies and eigenvectors. Converged results are obtained using a coarse $4\times4\times4$ $\mathbf{q}$-point grid which is used as a starting point for the Fourier interpolation to a finer grid along high symmetry lines to construct the phonon dispersion.

The phonon dispersions obtained with the LDA and HSE functionals are shown in Fig.~\ref{fig:phonon} without considering LO-TO splitting, and are in good agreement with ealier calculations.~\cite{bso_hse_bands_and_phonons} We estimate the effects of LO-TO splitting by calculating the phonon frequencies using elongated cells along the $(100)$ crystallographic direction with up to $32$ primitive cells within LDA, and up to $8$ primitive cells in HSE. The estimated LO-TO splittings are shown in Table~\ref{tab:lo-to} for the three infrared active $\Gamma$-point modes of $\Gamma_4^{-}$ symmetry, together with the corresponding experimental data from Ref.~\onlinecite{bso_phonons_bands}. The difference between the LDA results from the $(\frac{1}{8},0,0)$ and $(\frac{1}{32},0,0)$ $\mathbf{q}$-points provides an estimate of the error in the HSE results evaluated only at the $(\frac{1}{8},0,0)$ $\mathbf{q}$-point, and this error is in the submeV energy range for the two highest frequency modes, and in the meV range for the lowest frequency mode (see Table~\ref{tab:lo-to}). 

\begin{table*}
  \setlength{\tabcolsep}{2pt} 
  \caption{LO-TO splitting for BaSnO$_3$ using the LDA and HSE exchange correlation functionals, and corresponding experimental data from Ref.~\onlinecite{bso_phonons_bands}, for the infrared active $\Gamma_4^{-}$ modes. We show results for LDA estimated from the $\mathbf{q}$-points $(\frac{1}{8},0,0)$ and $(\frac{1}{32},0,0)$, and for HSE estimated from the $\mathbf{q}$-point $(\frac{1}{8},0,0)$. The optical modes considered have transverse frequencies of $\omega_1^{\mathrm{TO}}=17.4$\,meV, $\omega_2^{\mathrm{TO}}=29.5$\,meV, and $\omega_3^{\mathrm{TO}}=82.9$\,meV at the LDA level, and of $\omega_1^{\mathrm{TO}}=17.3$\,meV, $\omega_2^{\mathrm{TO}}=32.3$\,meV, and $\omega_3^{\mathrm{TO}}=80.6$\,meV at the HSE level.} 
  \label{tab:lo-to}
  \begin{ruledtabular}
  \begin{tabular}{c|cccc}
            & $\left(\frac{1}{8},0,0\right)$ LDA  & $\left(\frac{1}{32},0,0\right)$ LDA & $(\frac{1}{8},0,0)$ HSE & Experiment (Ref.~\onlinecite{bso_phonons_bands}) \\ [0.1cm]
  \hline
  $\omega_1^{\mathrm{LO-TO}}$ &  $3.1$\,meV  & $1.6$\,meV  & $3.1$\,meV  & $2.4$\,meV \\
  $\omega_2^{\mathrm{LO-TO}}$ &  $18.2$\,meV & $18.0$\,meV & $19.7$\,meV  &  $21.9$\,meV  \\
  $\omega_3^{\mathrm{LO-TO}}$ &  $10.0$\,meV & $10.3$\,meV & $11.0$\,meV  &  $11.8$\,meV \\
\end{tabular}
\end{ruledtabular}
\end{table*}

The comparison of the phonon frequencies between LDA and HSE shown in Fig.~\ref{fig:phonon} and Table~\ref{tab:lo-to} shows significant differences between the two. First, for the low-energy modes below about $55$\,meV, the LDA modes are in general softer than the corresponding HSE modes. The situation is reversed for the high energy modes, where the HSE frequencies are smaller. Second, the LO-TO splitting for the two higher energy modes is about $10$\% larger using HSE over LDA. The stronger LO-TO splitting in HSE is in closer agreement with the experimental measurements of Ref.~\onlinecite{bso_phonons_bands}. Third, the LDA results exhibit a triply degenerate soft mode at the $R$ point labeled by the irreducible representation $R_5^{-}$ (with the Ba atom at the origin of coordinates) and with a vibrational frequency of only $1.8$~meV. This mode is significantly harder at the HSE level of theory, reaching $11.3$~meV. Using the PBEsol or PBE functionals, the $R_5^{-}$ mode becomes imaginary, as shown in Table~\ref{tab:soft_mode}. As the cubic perovskite structure of BaSnO$_3$ is dynamically stable experimentally, our results suggest that the hybrid HSE functional provides a better description than semilocal functionals of the lattice dynamics of this system. This observation could have important implications for the study of superlattices formed by BaSnO$_3$, and in particular about their dynamical stability and ground state structures.

\section{Phonon-assisted optical absorption} \label{sec:optics}

\subsection{Formalism}

The optical constants of a solid can be derived from the complex dielectric function $\varepsilon_1+i\varepsilon_2$. In this paper, we describe the frequency dependent dielectric function within the dipole approximation, and write
\begin{equation}
\varepsilon_2(\omega)=\frac{2\pi}{mN}\frac{\omega^2_{\mathrm{P}}}{\omega^2}\sum_{v,c}\int_{\mathrm{BZ}}\frac{d\mathbf{k}}{(2\pi)^3}|M_{cv\mathbf{k}}|^2\delta(\epsilon_{c\mathbf{k}}-\epsilon_{v\mathbf{k}}-\hbar\omega), \label{eq:optical}
\end{equation}
where $m$ is the electron mass, $N$ is the number of electrons per unit volume, and $\omega^2_{\mathrm{P}}=4\pi Ne^2/m$ is the plasma frequency with $e$ the electron charge. The single-particle electronic states $|\psi\rangle$ of energy $\epsilon$ are labeled by their crystal momentum $\mathbf{k}$ and their valence $v$ or conduction $c$ band index. The sum is over connecting valence and conduction states, and over all $\mathbf{k}$-points in the Brillouin zone (BZ). The optical matrix element is given by $M_{cv\mathbf{k}}=\langle\psi_{c\mathbf{k}}|\hat{\mathbf{e}}\cdot\mathbf{p}|\psi_{v\mathbf{k}}\rangle$, where $\hat{\mathbf{e}}$ is the polarization of the incident light and $\mathbf{p}$ is the momentum operator. The real part of the dielectric function $\varepsilon_1(\omega)$ can be obtained from the imaginary part using the Kramers-Kronig relation.~\cite{ziman_solids} From the dielectric function, we calculate the absorption coefficient as $\kappa(\omega)=\omega\varepsilon_2(\omega)/cn(\omega)$, where $c$ is the speed of light and $n(\omega)$ is the refractive index.

Within the theory of Williams and Lax,~\cite{williams_phonon_assisted_optics,lax_phonon_assisted_optics} the imaginary part of the dielectric function at temperature $T$ is given by
\begin{equation}
\varepsilon_2(\omega;T)=\frac{1}{\mathcal{Z}}\sum_{\mathbf{s}}\langle\Phi_{\mathbf{s}}(\mathbf{u})|\varepsilon_2(\omega;\mathbf{u})|\Phi_{\mathbf{s}}(\mathbf{u})\rangle e^{-E_{\mathbf{s}}/k_{\mathrm{B}}T}, \label{eq:imag_tdep}
\end{equation}
where the harmonic vibrational wave function $|\Phi_{\mathbf{s}}(\mathbf{u})\rangle$ in state $\mathbf{s}$ has energy $E_{\mathbf{s}}$, $\mathbf{u}=\{u_{\nu\mathbf{q}}\}$ is a collective coordinate for all the nuclei written in terms of normal modes of vibration $(\nu,\mathbf{q})$, $\mathcal{Z}=\sum_{\mathbf{s}}e^{-E_{\mathbf{s}}/k_{\mathrm{B}}T}$ is the partition function, $T$ is the temperature, and $k_{\mathrm{B}}$ is Boltzmann's constant. Zacharias, Patrick, and Giustino~\cite{patrick_elph_long,phonon_assisted_abs_stochastic} established that the Williams-Lax expression in Eq.~(\ref{eq:imag_tdep}) is an adiabatic approximation to the standard expression for the temperature dependent imaginary part of the dielectric function within the theory of phonon-assisted optical absorption of Hall, Bardeen, and Blatt,~\cite{phonon_assisted_abs_hbb} and is valid as long as phonon energies are small compared to the size of the band gap, $\hbar\omega_{\nu\mathbf{q}}\ll\epsilon_c-\epsilon_v$. An advantage of the Williams-Lax theory is that the temperature dependence of the electronic band structure~\cite{0022-3719-9-12-013,PhysRevLett.105.265501,monserrat_elph_diamond_silicon,gonze_marini_elph,gonze_gw_elph} is automatically incorporated,~\cite{giustino_nat_comm,patrick_elph_long,phonon_assisted_abs_stochastic,phonon_assisted_abs_one_shot} while within the Hall-Bardeen-Blatt theory the band structure is temperature-independent. In this work, we use Eq.~(\ref{eq:imag_tdep}) to study phonon-assisted optical absorption in BaSnO$_3$.

We evaluate Eq.~(\ref{eq:imag_tdep}) using thermal lines (TL) as introduced in Ref.~\onlinecite{thermal_lines}. In this approach, the multidimensional integral over the harmonic vibrational density is evaluated using the mean value (MV) theorem for integrals. This theorem dictates that there exists at least one atomic configuration, which we denote by $\mathbf{u}^{\mathrm{MV}}(T)$ with an explicit temperature dependence $T$, for which $\varepsilon_2(\omega;\mathbf{u}^{\mathrm{MV}}(T))=\varepsilon_2(\omega;T)$. This would, in principle, allow us to replace the multidimensional integral in Eq.~(\ref{eq:imag_tdep}) by the evaluation of the integrand on a single atomic configuration $\mathbf{u}^{\mathrm{MV}}(T)$ at each temperature of interest. Following Ref.~\onlinecite{thermal_lines}, we can find a good approximation to $\mathbf{u}^{\mathrm{MV}}(T)$ by choosing an atomic configuration for which each phonon mode $(\nu,\mathbf{q})$ has an amplitude given by
\begin{equation}
u^{\mathrm{TL}}_{\nu\mathbf{q}}(T)=\pm\left(\frac{1}{\omega_{\nu\mathbf{q}}}\left[\frac{1}{2}+n_{\mathrm{B}}(\omega_{\nu\mathbf{q}},T)\right]\right)^{1/2}, \label{eq:tl}
\end{equation}
where $\omega_{\nu,\mathbf{q}}$ is the phonon frequency, and $n_{\mathrm{B}}$ is a Bose-Einstein factor. We note that the expression in Eq.~(\ref{eq:tl}) is such that for each phonon mode there are two possible amplitudes, thus the number of mean value points is $2^{3(\mathcal{N}-1)}$, where $\mathcal{N}$ is the number of atoms in the system.  Configurations on thermal lines are the exact mean value configurations if the integrand $\varepsilon_2(\omega;\mathbf{u})$ is a quadratic function of ${u_{\nu\mathbf{q}}}$. In practise, we stochastically sample a subset of configurations on thermal lines, and find that the finite temperature dielectric function converges using only two sampling points. We refer the reader to Refs.~\onlinecite{thermal_lines,gw_thermal_lines} for further details about thermal lines.

\subsection{Computational details}

Our first principles calculations of the imaginary part of the dielectric function $\varepsilon_2(\omega;\mathbf{u})$ are performed using {\sc vasp}. We report results obtained using a $350$\,eV energy cut-off, and we sample the electronic BZ stochastically including $6400$ $\mathbf{k}$-points (equivalent to $100$ $\mathbf{k}$-points on a $4\times4\times4$ supercell). The energy conservation for the optical absorption process imposed by the delta function in Eq.~(\ref{eq:optical}) is smeared with a Gaussian function of width $80$\,meV. We report results using both the LDA and HSE exchange correlation functionals.

For the electron-phonon contribution to optical absorption we report results obtained using a $4\times4\times4$ supercell of BaSnO$_3$ containing $320$ atoms, which is equivalent to sampling the vibrational BZ using a $4\times4\times4$ $\mathbf{q}$-point grid. Tests with a $5\times5\times5$ supercell show small variations on the optical spectrum, but these do not affect our conclusions. We find that in the evaluation of Eq.~(\ref{eq:imag_tdep}) using thermal lines, a single configuration is sufficient to obtain converged results, which suggests that the dependence of $\varepsilon_2$ on the phonon modes is close to quadratic. The reported results are obtained averaging over two configurations.

Full convergence tests are detailed in the Supplemental Material.

\subsection{Absorption onset}

\begin{figure}
\centering
\includegraphics[scale=0.35]{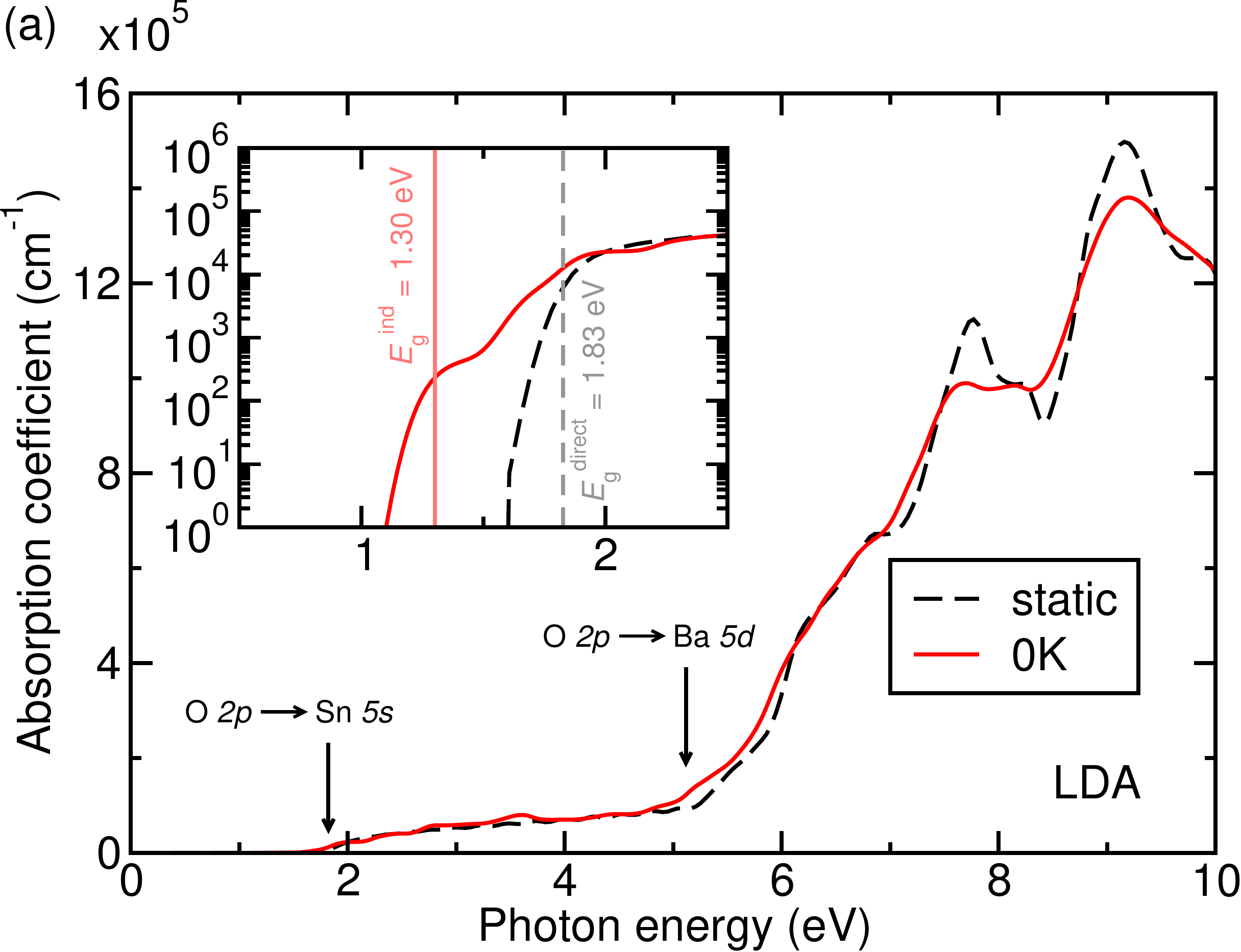}
\includegraphics[scale=0.35]{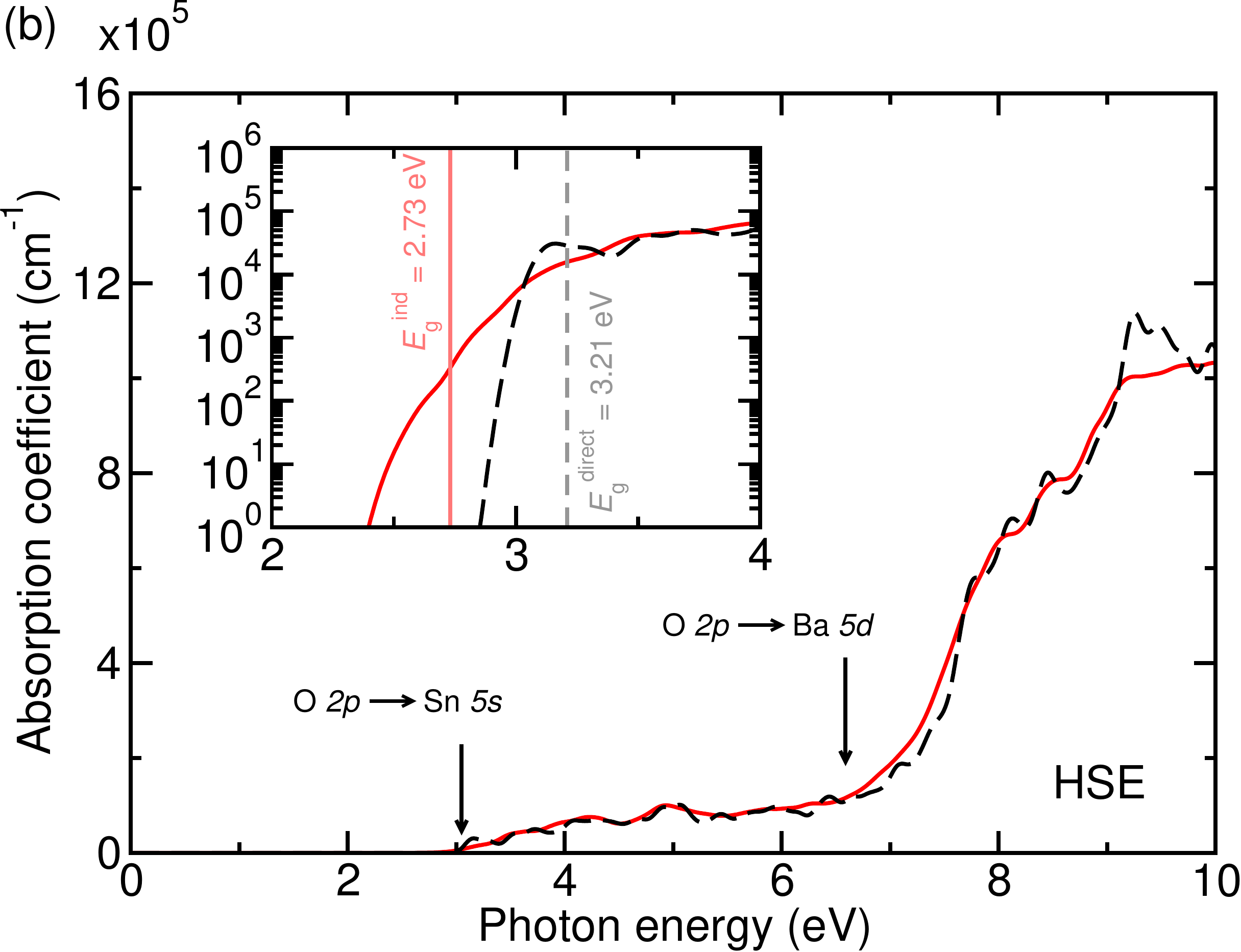}
\caption{Absorption coefficient of BaSnO$_3$ using (a) the LDA and (b) the HSE approximations to the exchange correlation functional. Calculations correspond to the inclusion of phonon-assisted contributions at $0$~K (solid red lines) and to stationary atoms at their equilibrium positions (dashed black lines). The inset depicts the same data near the absorption onset, with the vertical lines indicating the positions of the minimum indirect and direct gaps. We note that the absorption onset occurs below the nominal direct band gap due to the smearing of the delta function in Eq.~(\ref{eq:optical}), as discussed in the text.} 
\label{fig:absorption}
\end{figure}

The absorption spectrum of BaSnO$_3$ is shown in Fig.~\ref{fig:absorption}. Focusing on the static lattice results first (dashed black lines), which correspond to vertical optical transitions only, both LDA and HSE calculations exhibit the same features. The absorption onset occurs at the minimum direct gap, located at the $\Gamma$ point between an O $2p$ valence state and a Sn $5s$ conduction state. The static lattice absorption onset occurs around $1.8$\,eV at the LDA level, and about $3.2$\,eV at the HSE level. For a range of about $4$\,eV, absorption only occurs between the valence O $2p$ states and the isolated Sn $5s$ conduction band (cf. Fig.~\ref{fig:bs}). Between $5$ and $7$\,eV there is a dramatic increase in absorption, determined by the energy of the transitions between the valence O $2p$ states and the conduction Ba $5d$ states. This increase in absorption occurs around $5.3$\,eV in LDA and $7.0$\,eV in HSE. We note that the absorption onset occurs below the nominal direct band gap due to the smearing by $80$\,meV of the delta function in Eq.~(\ref{eq:optical}), with the logarithmic scale used in the insets making the apparent shift larger. The tail below the nominal band gap that we observe is similar to that observed in earlier calculations in silicon.~\cite{phonon_assisted_abs_stochastic}

We next consider the results at $0$\,K shown as the red solid lines in Fig.~\ref{fig:absorption} and obtained including the effects of electron-phonon coupling. These results differ from the static lattice results because they include the effects of quantum zero-point motion, which in the perturbative point of view correspond to phonon emission, and enable indirect transitions to occur. The most important difference between the static lattice results and the zero temperature results is the nature of the absorption onset. The minimum absorption onset at $0$\,K is determined by a second order process which involves the absorption of a photon and the scattering off a phonon. This process bridges the indirect gap of $1.30$\,eV (LDA) or $2.73$\,eV (HSE) and can only be accounted for theoretically by the use of the theory of phonon-assisted optical absorption. In both LDA and HSE calculations, the zero temperature results demonstrate an absorption onset that is about $0.5$\,eV smaller than predicted by the static lattice theory.

\subsection{Temperature dependent absorption}

\begin{figure}
\centering
\includegraphics[scale=0.35]{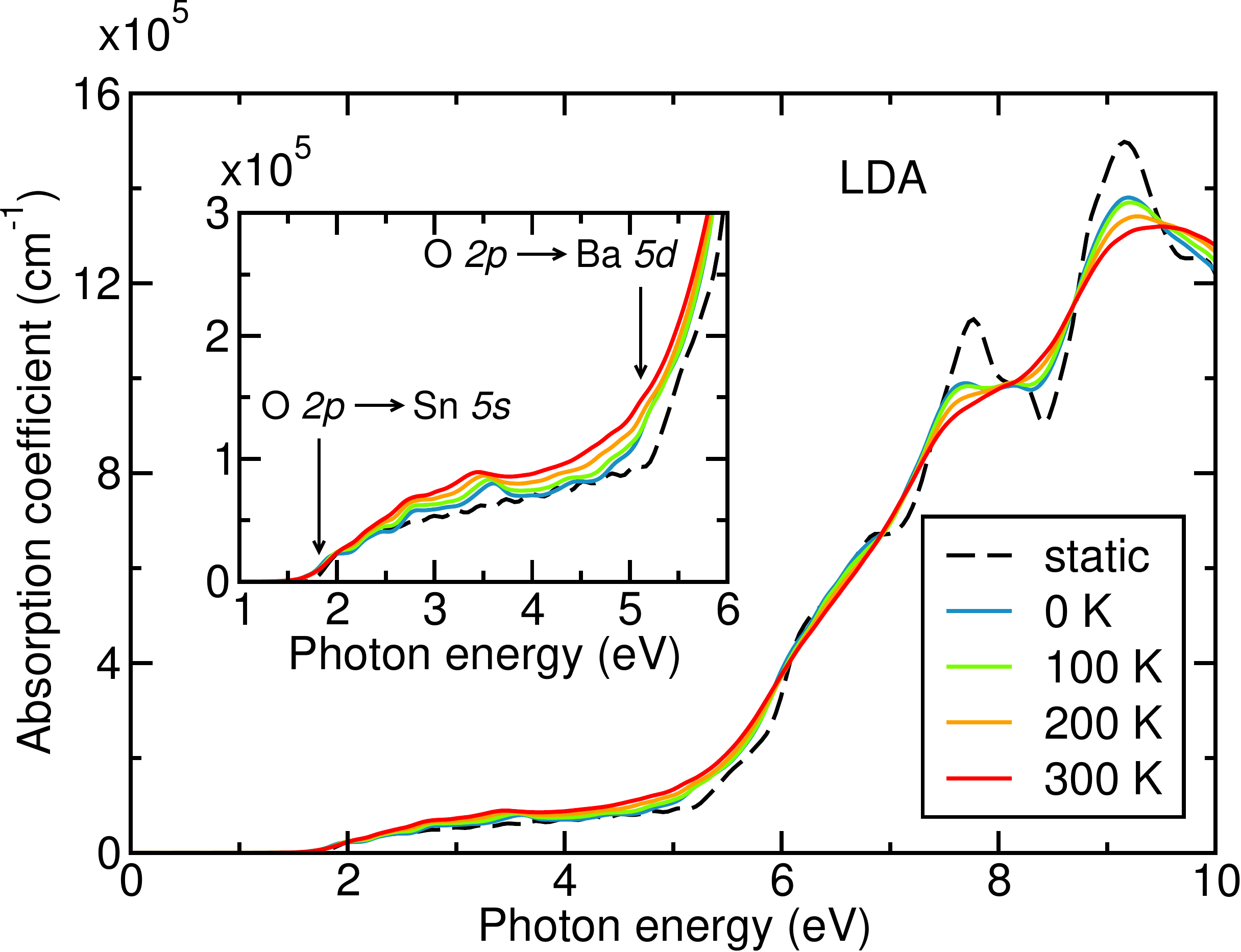}
\caption{Temperature dependence of the absorption coefficient of BaSnO$_3$ using the LDA approximation to the exchange correlation functional. The inset details the absorption coefficient in the energy range in which the conduction Sn $5s$ band dominates.}
\label{fig:absorption-temp}
\end{figure}

In Fig.~\ref{fig:absorption-temp} we show the temperature dependent absorption spectrum of BaSnO$_3$ at the LDA level of theory. The first noteworthy feature of the finite temperature absorption spectrum is the smoothing of the peaks exhibited by the static lattice spectrum. As an example, the static lattice absorption peak just below $8$\,eV decreases in size with increasing temperature, and becomes a shoulder of the larger peak around $9$\,eV. 

A second feature of the finite temperature results is the increase in the absorption coefficient in the energy range from about $2$ to $5$\,eV as temperature increases (shown in the inset of Fig.~\ref{fig:absorption-temp}). This energy range corresponds to transitions from the valence O $2p$ bands to the isolated conduction Sn $5s$ band (cf. Fig.~\ref{fig:bs}). To understand the origin of this effect, recall that at the static lattice level only vertical transitions are allowed. This implies that in each energy interval there are only a small number of conduction states available for electronic transitions (those of the Sn $5s$ band). Furthermore, for each of these conduction states there is only a small number of O $2p$ states from which electrons can absorb photons, those at the same $\mathbf{k}$-vector as the conduction band. This explains why the absorption coefficient is small below $5$\,eV when only transition to the Sn $5s$ band are allowed. When the effects of electron-phonon coupling are included, then the phase space of available electronic states on the conduction band remains the same, but electrons from generic $\mathbf{k}$-points in the flat O $2p$ bands can now absorb a photon with the mediation of a phonon. This is a second order process, and therefore its weight is smaller than the dominant vertical first order absorption process. But with increasing temperature the phonon-mediated processes become more relevant, and this is reflected in the increase in the absorption coefficient at energies between $2$ and $5$\,eV.

A final feature of the finite temperature results is the red shift of the absorption onset of the O~$2p\rightarrow$~Ba~$5d$ transition as temperature increases, which is clearly observed in the inset of Fig.~\ref{fig:absorption-temp}. 

\begin{figure}
\centering
\includegraphics[scale=0.45]{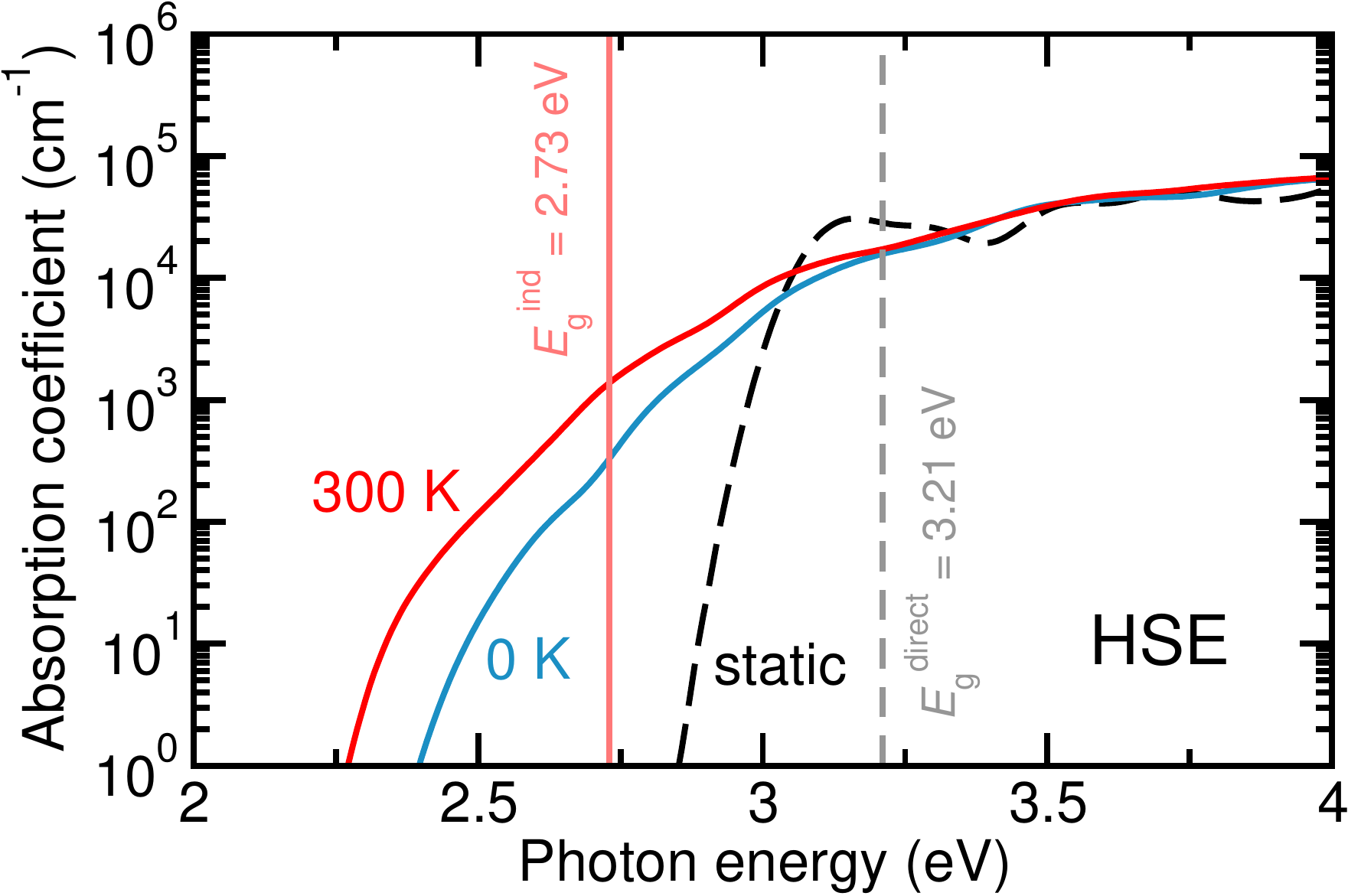}
\caption{Absorption coefficient of BaSnO$_3$ at the static lattice level, $0$\,K, and $300$\,K using the HSE approximation to the exchange correlation functional.} 
\label{fig:absorption-temp-hse}
\end{figure}

We expect that qualitatively similar finite temperature features would be observed if the HSE functional was used instead. However, the computational expense of the latter precludes detailed calculations, and we have performed finite temperature calculations for HSE only at $300$\,K, shown in Fig.~\ref{fig:absorption-temp-hse}. The most clear feature is a red shift in the indirect absorption onset between $0$\,K and $300$\,K of about $0.13$\,eV. The corresponding red shift of the indirect absorption onset at the LDA level is about $0.03$\,eV. This observation is in line with earlier reports of stronger electron-phonon coupling when electronic correlations beyond semilocal DFT are included.~\cite{gonze_gw_elph,gw_thermal_lines} 


\subsection{Discussion}

Our first principles calculations provide the first study of the optical absorption spectrum of BaSnO$_3$ including the effects of electron-phonon coupling and electron-electron interactions beyond semilocal DFT. Our results allow us to confirm that the onset of optical absorption in BaSnO$_3$ is due to the indirect gap, as previously suggested from experimental absorption spectra.~\cite{bso_exp_characterization} We expect that these calculations will become more widespread in the future and phonon-assisted processes will no longer be limited to experimental analysis but will also be routinely treated at the theoretical level.

We further provide results of the temperature dependence of the absorption spectrum of BaSnO$_3$. The calculated red shift of the absorption onset with increasing temperature is small in BaSnO$_3$, $0.1$\,eV from $0$\,K to $300$\,K, suggesting that the performance of transparent conductors based on BaSnO$_3$ will only weakly depend on temperature. The increase in magnitude of the absorption coefficient in the energy range dominated by the lone Sn $5s$ band might be a generic feature of transparent conducting oxides because their high conductivities rely on the same principle of an isolated $s$-character conduction band. 

More generally, it would be interesting to explore the implications of phonon-assisted processes in the novel design strategies that have been proposed for transparent conductors beyond oxide semiconductors.~\cite{tc_ternary_zunger,tco_graphene,tco_metal_nanowires,tco_correlated_metal,zunger_bulk_band_engineering} In these proposals transparency is typically achieved by recourse to selection rules or momentum mismatch between band extrema that forbid some optical transitions within the static lattice approximation. Phonons can both break symmetry-based selection rules and allow phonon-assisted transitions of finite momentum, which might invalidate some of the proposed strategies. However, these phonon-assisted processes are second-order, and a quantitative assessment of their importance is required before any definite conclusions can be reached.


\section{Summary and conclusions} \label{sec:conclusions}

We have presented first principles calculations of the absorption spectrum of the transparent conducting oxide BaSnO$_3$, including both electron-phonon coupling and electron-electron coupling beyond semilocal density functional theory. Our results demonstrate that both effects are necessary in order to obtain a qualitatively and quantitatively accurate spectrum. Electron-phonon coupling permits phonon-assisted optical absorption across the minimum indirect gap of BaSnO$_3$, a transition that is forbidden in the standard static lattic approximation. This provides an absorption onset that occurs about $0.5$\,eV below the previously calculated direct absorption onset. Electron-phonon coupling also leads to the temperature dependence of the absorption spectrum of BaSnO$_3$. Electron-electron correlations treated at the hybrid functional level of theory indicate that the conduction bands span a wider range of energies than predicted by semilocal functionals, and therefore modify the position of the absorption peaks in a manner that cannot be predicted by a simple rigid shift of the bands.

Our work demonstrates that an accurate description of the optical properties of BaSnO$_3$ requires the inclusion of both electron-phonon and electron-electron terms. Recent methodological developments make the inclusion of these terms feasible in the context of first principles calculations, and we think that the prediction of novel materials for optoelectronic applications will benefit from these highly accurate calculations that model experimental settings more closely than standard approaches.

\acknowledgments

The authors thank Heung-Sik Kim and Andr\'{e} Schleife for helpful discussions and correspondence. This work was partially supported by NSF grant DMR-1629346. B.M. thanks Robinson College, Cambridge, and the Cambridge Philosophical Society for a Henslow Research Fellowship. 

\bibliography{bso}

\onecolumngrid
\clearpage
\begin{center}
\textbf{\large Supplemental Material for ``Phonon-assisted optical absorption in BaSnO$_3$ from first principles''}
\end{center}
\setcounter{equation}{0}
\setcounter{figure}{0}
\setcounter{table}{0}
\setcounter{page}{1}
\makeatletter
\renewcommand{\theequation}{S\arabic{equation}}
\renewcommand{\thefigure}{S\arabic{figure}}
\renewcommand{\bibnumfmt}[1]{[S#1]}
\renewcommand{\citenumfont}[1]{S#1}

In the Supplemental Material we evaluate the various convergence parameters of the calculation of the finite temperature absorption coefficient of BaSnO$_3$. For the convergence study, we use the local density approximation to the exchange correlation functional~\cite{PhysRevLett.45.566,PhysRevB.23.5048,PhysRevB.45.13244}, and perform all calculations at $300$\,K.

\begin{figure}[h]
\includegraphics[width=0.4\textwidth]{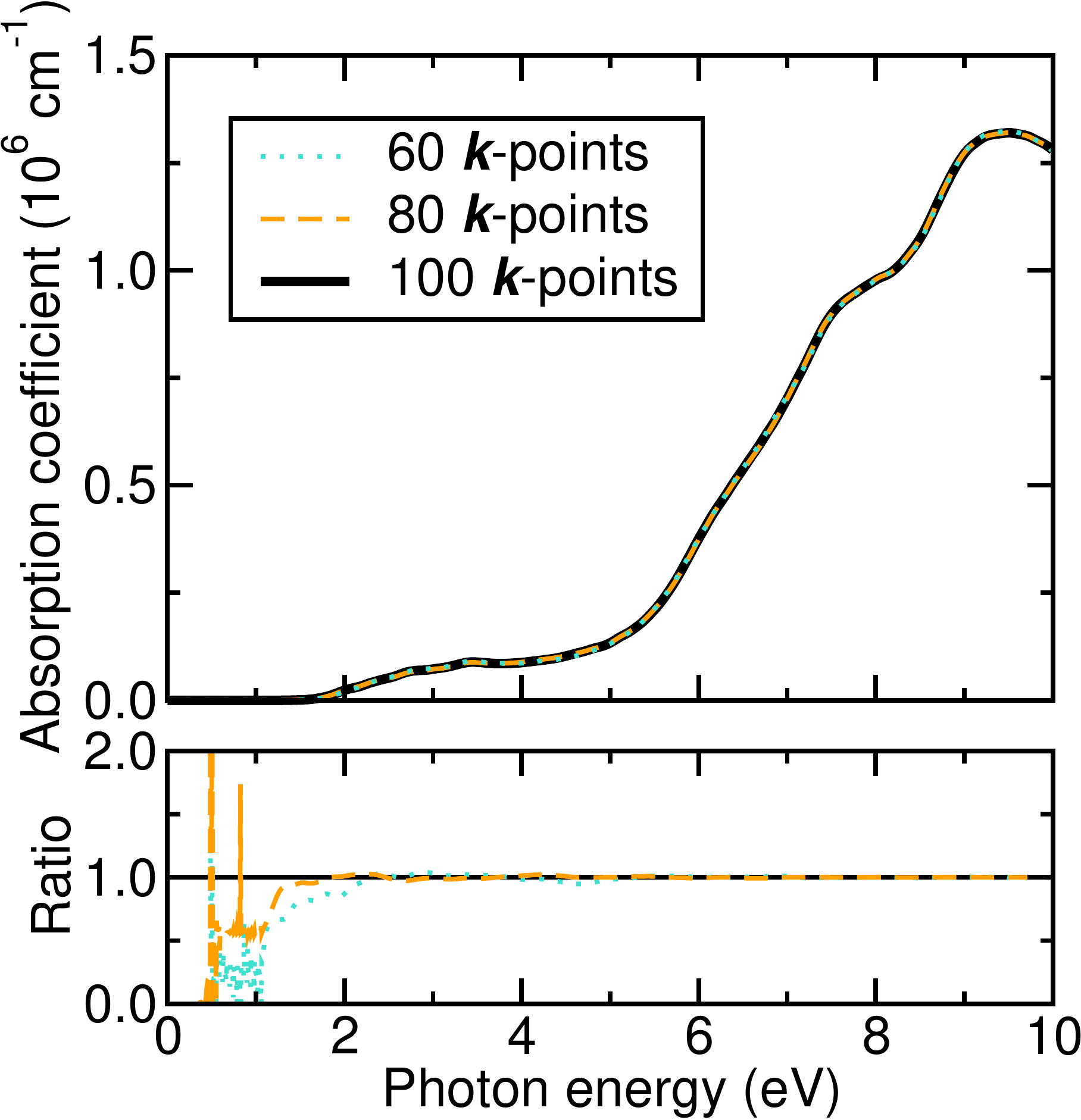}
\hspace{0.7cm}
\includegraphics[width=0.4\textwidth]{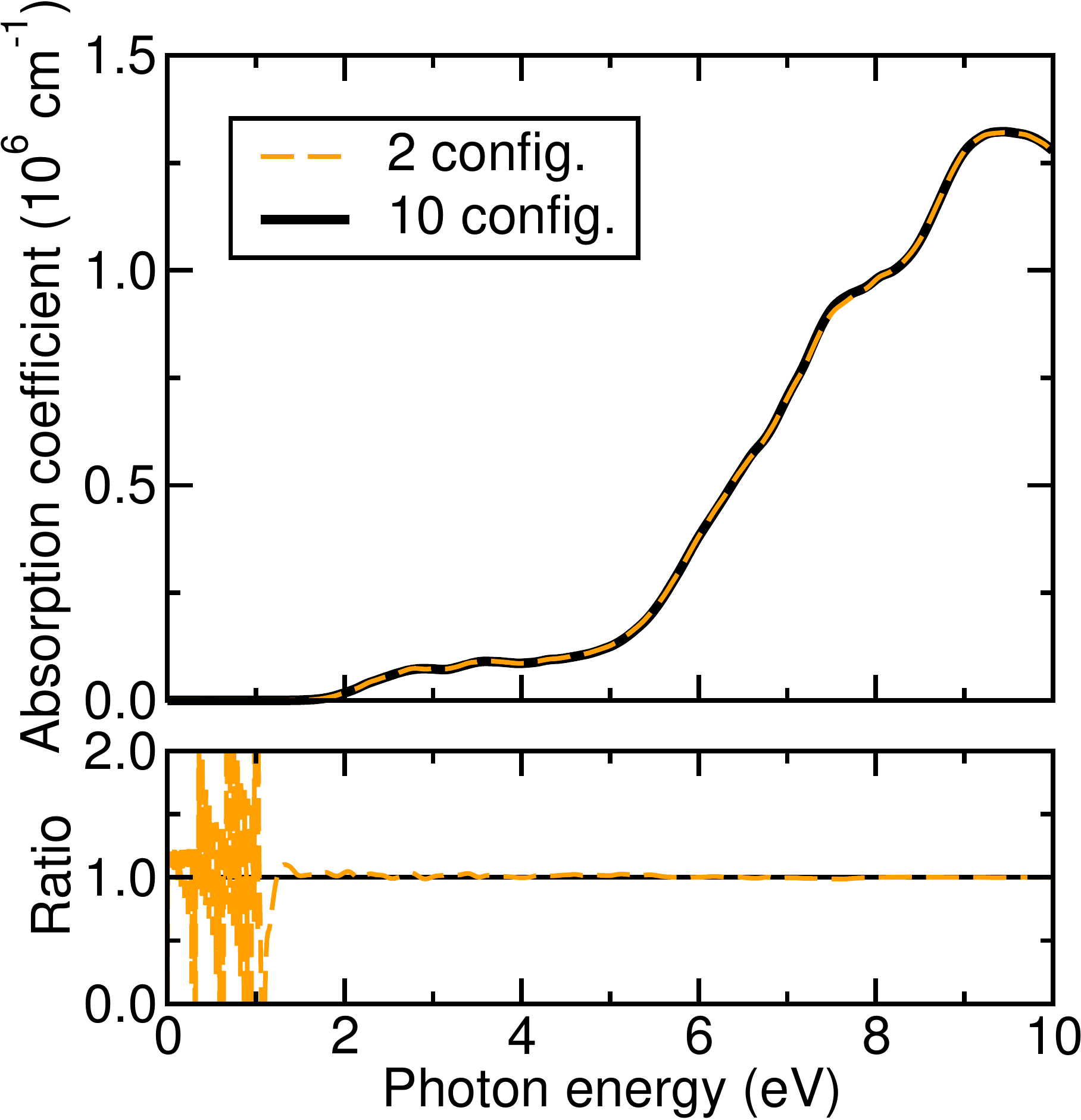}
\caption{Absorption coefficient of BaSnO$_3$ at $300$~K calculated using a $4\times4\times4$ supercell. The left diagram depicts convergence with respect to the number of random electronic $\mathbf{k}$-points used to sample the electronic Brillouin zone. The right diagram depicts the convergence with respect to the number of thermal line configurations used in the thermodynamic average.}
\label{fig:kpoints}
\end{figure}

In Fig.~\ref{fig:kpoints} we show calculations corresponding to a $4\times4\times4$ supercell. On the left diagram of Fig.~\ref{fig:kpoints} we show the absorption coefficient calculated using $60$, $80$, and $100$ random $\mathbf{k}$-points to sample the electronic Brillouin zone (BZ). These numbers correspond to $3840$, $5120$, and $6400$ $\mathbf{k}$-points, respectively, in the BZ of the primitive cell. The bottom pannel depicts the ratio of the absorption coefficients with respect to the curve with $100$ random $\mathbf{k}$-points, demonstrating the convergence with respect to the number of $\mathbf{k}$-points used to sample the electronic BZ. For photon energies below about $1$\,eV, the ratio exhibits random oscillations which are caused by numerical noise. The value of the absorption coefficient at energies below about $1$\,eV is smaller than $1$\,cm$^{-1}$, the lowest absorption coefficient reported in the main mansucript.

On the right diagram of Fig.~\ref{fig:kpoints} we show the absorption coefficient calculated using different numbers of atomic configurations to sample the vibrational phase space, all configurations corresponding to a thermal line~\cite{thermal_lines}. Using only $2$ configurations leads to converged results as shown both in the upper pannel for the absorption coefficient, and in the lower pannel for the ratio. We again observe random oscillations due to numerical noise for photon energies below about $1$\,eV.

\begin{figure}
\includegraphics[width=0.4\textwidth]{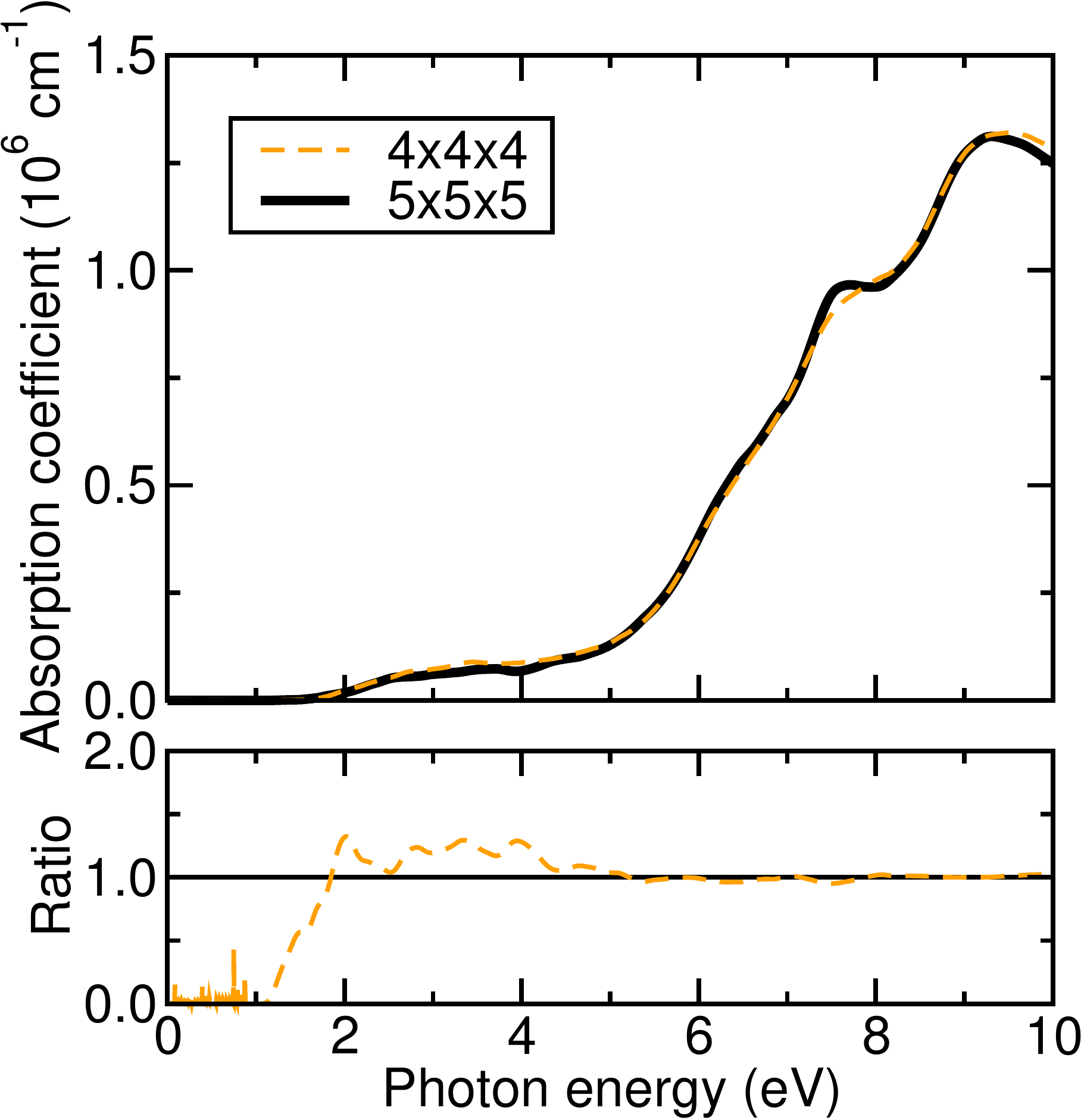}
\caption{Absorption coefficient of BaSnO$_3$ at $300$~K calculated using a $4\times4\times4$ and a $5\times5\times5$ supercell.}
\label{fig:sc}
\end{figure}

The sampling of the vibrational BZ is accomplished by the use of supercells of the $5$-atom primitive cell of BaSnO$_3$. In Fig.~\ref{fig:sc} we compare the absorption coefficient obtained using supercells of sizes $4\times4\times4$ (containing $320$ atoms) and $5\times5\times5$ (containing $625$ atoms). The results show some marked differences between the two calculations, in particular the absorption peak just below $8$\,eV is slightly stronger for the larger $5\times5\times5$ supercell, and the absorption coefficient in the range between $2$\,eV and $5$\,eV is slightly larger for the smaller $4\times4\times4$ supercell. Nonetheless, the results presented in the main text are robust with respect to the size of the supercell, and therefore we use a $4\times4\times4$ supercell in our calculations as it delivers the appropriate balance between accuracy and computational cost.

We finally note that for the calculation of the imaginary part of the dielectric function, energy conservation is imposed by smearing the delta function with a Gaussian. The results reported in the main text correspond to using a smearing width of $80$\,meV. Tests with a smearing width of $20$\,meV show that our conclusions are independent of the smearing width used.

Overall, the results reported in the main text are obtained using a $4\times4\times4$ supercell, with averaging over two configurations on thermal lines, and including $100$ random $\mathbf{k}$-points in the supercell (equivalent to $6400$ $\mathbf{k}$-points in the primitive cell).

\end{document}